\documentclass{INTERSPEECH2023}
\usepackage{float}

\interspeechcameraready

\title{Towards Robust Family-Infant Audio Analysis Based on Unsupervised Pretraining of Wav2vec 2.0 on Large-Scale Unlabeled Family Audio}
\name{Jialu Li$^{1,2}$, Mark Hasegawa-Johnson$^{1,2}$, Nancy L. McElwain$^{2,3}$}
\address{
  $^1$Department of Electrical and Computer Engineering, University of Illinois\\ 
  $^2$Beckman Institute for Advanced Science and Technology, University of Illinois \\
  $^3$Department of Human Development and Family Studies, University of Illinois}
\email{jialuli3@illinois.edu, jhasegaw@illinois.edu, mcelwn@illinois.edu}

\begin{document}

\maketitle
 
\begin{abstract}
To perform automatic family audio analysis, past studies have collected recordings using phone, video, or audio-only recording devices like LENA, investigated supervised learning methods, and used or fine-tuned general-purpose embeddings learned from large pretrained models.
In this study, we advance the audio component of a new infant wearable multi-modal device called LittleBeats (LB) by learning family audio representation via wav2vec 2.0 (W2V2) pretraining. 
We show given a limited number of labeled LB home recordings, W2V2 pretrained using 1k-hour of unlabeled home recordings outperforms oracle W2V2 pretrained on 960-hour unlabeled LibriSpeech in terms of parent/infant speaker diarization (SD) and vocalization classifications (VC) at home. Extra relevant external unlabeled and labeled data further benefit W2V2 pretraining and fine-tuning.
With SpecAug and environmental speech corruptions, we obtain 12\% relative gain on SD and moderate boost on VC. 
Code and model weights are available.

\end{abstract}

\noindent\textbf{Index Terms}: family-infant audio analysis, speaker diarization, vocalization classification, unsupervised learning, wav2vec 2.0
\let\thefootnote\relax\footnotetext{This work was supported by funding from NIDA (R34DA050256), NIMH (R21MH112578), NIFA (ILLU-793-368), NSF (1725729), and Beckman Graduate Fellowship.}

\vspace{-0.3cm}
\section{Introduction}
\begin{table*}
  \centering
  \setlength{\tabcolsep}{2.0pt}
   \caption{\small Overview of data distribution prepared after preprocessing steps. Last six columns are Total duration (including silence), Vocalized duration (all participants), and vocalized duration broken down into (CRY/FUS/BAB) for key child (CHN), (CDS/ADS/LAU/SNG) for adult female (FAN) and male (MAN), and total non-key child/sibling (CXN) in hours (h). In:in-domain/Out:out-of-domain. $m$:month/$y$:year.}
    \label{tab:data}
      \vspace{-0.2cm}
\begin{adjustbox}{max width=1\textwidth}
  \begin{tabular}{cccccc|cccccc}
    \toprule
    Device & Type & Domain & Context & Age & \# of families & Total dur &Vocal dur & CHN & FAN & MAN & CXN  \\
    \midrule
    LB & Labeled & In & Home & $<14m$ & 22 & 10.61h & 4.78h & .18/.48/1.1 & 1.0/.68/.04/.11 & .24/.34/.01/.03 & .57 \\ 
    LENA & Labeled & Out & Home & $<24m$ & 30 & 14.59h & 9.05h & .58/.75/.84 & 1.11/1.97/.08/.50 & 1.24/.82/.05/.09 & 1.02  \\
    LB & Labeled & Out & Virtual visit & $<10m$ & 11 & 1.35h & 0.8h & .02/.14/.14 & .38/.03/.01/.08 & - & -\\
    Camera & Labeled & Out & Lab visit & $<14m$ & 105 & 9.94h & 3.4h & .24/.69/.23 & 1.75/.01/.1/.38 & - & - \\ 
    \midrule
    LB & Unlabeled & In & Home & $<5y$ & 110 & 1100h & - & - & - & - & - \\
    LENA & Unlabeled & Out & Home & $<5y$ & 113 & 3200h & - & - & - & - & - \\
    \bottomrule
  \end{tabular}
  \end{adjustbox}
\label{tab:classify}
  \vspace{-0.5cm}
\end{table*} 

In the U.S., 1 in 6 children aged 2–8 years has a diagnosed mental, behavioral, or developmental disorder~\cite{bitsko2016health}, but such disorders are often neglected. Child mental health problems begin in early childhood, and daily interactions with family members that are repeated and reinforced over time are critical to children’s emotional well-being. According to attachment theory: the primary caregivers who respond quickly and consistently to an infant's needs allow the child to develop a sense of security. Children may develop insecure attachment styles if parents are often unavailable, intrusive, or respond inconsistently to the child's cues, particularly signs of distress~\cite{mcelwain2006maternal}. 
Additionally, previous psychological studies indicate that parents and infants are more likely to show coordinated physiological activities when their vocal and physical behaviors are mutually responsive
during play~\cite{hu2021mother}, and that such mutually responsive behaviors help maintain or increase shared positive interactions and emotions~\cite{kochanska2005pathways}. Therefore, to better support child mental health outcomes, it is essential to detect if parent and infant establish coordinated behaviors at an early stage in daily activities at home. Although previous work emphasizes the importance of the mother-infant dyad, father-infant~\cite{sethna2017father} and sibling-infant~\cite{oh2015trajectories} interactions also play a crucial role in infant development. Thus, we consider the larger family context in this work to better understand infant emotional and behavioral development during the first years of life.
In the past, to analyze family interactions, researchers or parents have recorded family audio at home or laboratory using a cell phone, video camera, or an audio-only recording device like the Language Environment Analysis device (LENA) \cite{LENA}. In this study, we test a new infant wearable multi-modal device we developed called LittleBeats$^{\text{TM}}$ (LB), and we aim to advance the LB audio pipeline such that it automatically provides reliable labels of SD and VC for family members, including infants, parents, and siblings, at home. 


For automatic family audio analysis, previous studies used supervised machine learning models on adult-child diarization~\cite{xie2019multi,cristia2018talker}, detecting infant cry~\cite{9591582,jian2021research}, and classifying parent/infant vocalizations~\cite{lavechin2020open} with a limited amount of labeled vocalization data because audio annotation is a time-consuming and labor-intensive task. To tackle challenges of data sparsity, past studies explored transfer learning techniques on infant cry detection by incorporating external relevant labeled datasets for additional training~\cite{gujral2019leveraging,li2021analysis}, computing node similarities in graphical convolutional networks using both labeled and unlabeled datasets~\cite{9449246}, and fine-tuning~\cite{xu2022differential} or leveraging embeddings~\cite{yao2022infant} from models pretrained on large-scale image or audio datasets.
Recently, a self-supervised learning model, W2V2~\cite{wav2vec}, which uses unsupervised pretraining from 960 hours of unlabeled adult speech, excels at multiple downstream speech processing tasks, including speech-to-text~\cite{schneider2019wav2vec}, speech emotion recognition~\cite{pepino2021emotion}, and speaker diarization~\cite{edselc.2-52.0-8514006320920220101}. Fine-tuning pretrained W2V2 general-purpose audio representations has shown robust performance in some audio-related healthcare applications with resource-poor data, such as detecting stuttering~\cite{sheikh2022introducing} and classifying COVID-19 vocalization~\cite{9926967}, but in other applications, such as autism detection in child speech~\cite{chi2022classifying} and pathological speech recognition~\cite{violeta2022investigating}, W2V2 did not always outperform supervised learning models potentially due to the domain mismatch between pretraining and fine-tuning.
Limited research has assessed pretraining effects of W2V2 to learn task-specific audio representation; 
one example~\cite{chen2021exploring} showed task-adaptive pretraining on speech emotion recognition to bridge the gap between W2V2 pretraining and a target domain by continuing pretraining on a target dataset after initial pretraining.

In this paper, we explore benefits of learning family audio representation in W2V2 pretraining using large-scale day-long home recordings collected from LB and LENA. To the best of our knowledge, this is the first study that examines the pretraining effects of a self-supervised model on a family audio analysis task. 
We demonstrate that given a limited amount of labeled LB home recordings, W2V2 pretrained using 1100 hours of unlabeled LB home recordings outperforms oracle W2V2 pretrained using 960 hours adult speech in the fine-tuned downstream tasks of SD and VC. We also discover that pretraining and fine-tuning steps can further benefit from an additional 3200 hours of LENA unlabeled home recordings and from relevant labeled LENA home and laboratory recordings, respectively. 
This work is well-aligned with the Interspeech 2023 theme of inclusive speech technology. 
The current standard method for diagnosing behavioral disorders is observation in the clinic.  The need for clinical observation can make it difficult for working parents, and for parents in rural settings, to obtain an accurate diagnosis.  At-home diagnosis supported by LB technology would better serve poor and rural populations, and thus increase the inclusiveness of healthcare.

\vspace{-0.3cm}
\section{Data}
For this study, we recruited families with study flyers distributed at multiple local community organizations and online family forums. All study procedures were approved by the Institutional Review Board at the University of Illinois at Urbana-Champaign (UIUC).
To encourage participants to permit the recording of their private family life, our consent form specifies that data will not be shared outside of our research team.  Our consent form further specifies that most of the recordings will be processed automatically without human intervention (the unlabeled data), and that human coders will only listen to small samples of the data (the labeled data).  Many families noted that they were willing to participate only because the vast majority of recordings are processed automatically without human auditing.
We aim to improve performance on LB home recordings, so we consider LB home recordings as in-domain data and other relevant data as out-of-domain data. Table \ref{tab:data} summarizes the distributions of data used in this experiment after preprocessing steps (see Section \ref{sec:preprocess}).

\vspace{-0.2cm}
\subsection{Unlabeled data}
We collected a large amount of unlabeled day-long home recordings from families with children under 5 years of age. Child participants wore either the LB or LENA device at home during the day for two or three days.
Families who participated in LB and LENA home recordings do not overlap. 

\vspace{-0.2cm}
\subsection{Labeled data}
\label{sec:labeled_data}
To manually annotate in-domain LB home recordings, we separated each daylong recording into 10-min segments. As continuous manual annotation of the audio recordings is time- and labor-intensive, human coders only annotated a few 10-min segments for each family, selected based on the highest active vocalization rates computed by a statistical voice activity detector (VAD). 
Human coders manually labeled key child (CHN), female adult (FAN), male adult (MAN), and other child/sibling (CXN) vocalizations using Praat~\cite{boersma2006praat}, with cross-coder validation at a precision of 0.2s. Ten percent of selected segments were double-coded, and inter-coder reliability (Cohen's kappa score) was 0.89 for CHN, 0.86 for FAN, 0.81 for MAN, and 0.80 for CXN. Child vocalizations were manually labeled as \textit{cry (CRY)}, \textit{fuss (FUS)}, and \textit{babble (BAB)}; adult vocalizations (MAN and FAN) were manually labeled as \textit{child-directed speech (CDS)}, \textit{adult-directed speech (ADS)}, \textit{laugh (LAU)}, and \textit{singing/rhythmic speech (SNG)}. The Cohen's kappas were 0.76 for CHN, 0.87 for FAN, and 0.71 for MAN. In total, we obtained 70 labeled 10-mins segments from 22 families. 

We also labeled out-of-domain datasets from two studies of infant (3-12 months) and toddler (18-24 months) development.  We followed similar data collection and annotation protocol for out-of-domain datasets. 
One of the studies included annotations of LENA home recordings.
Another study recorded mother and infant completing two semi-structured interactive tasks
in the laboratory (Lab visit)
or at home using LB during the COVID-19 pandemic (Virtual visit).
The kappas for out-of-domain data are similar to in-domain data. Families who
participated in in- and out-of-domain studies do not overlap. 
\vspace{-0.2cm}
\subsection{Data preprocessing}
\label{sec:preprocess}
LB audio was sampled at 15832Hz or 22756Hz at two hardware versions. Camera and LENA audio were sampled at 48k Hz and 16k Hz respectively. 
To make audio data compatible with W2V2 training, we resampled LB and camera audio to 16kHz using librosa (v0.9.2)~\cite{mcfee2015librosa}. To prepare unlabeled data for pretraining, we first applied VAD to remove silent portions for all unlabeled home recordings, and then divided non-silent audio into 10s segments. To prepare labeled data for fine-tuning, we labeled the audio stream in intervals of 2s starting every 0.2s. 
The label of each 2s interval was determined by the temporal majority of human annotations on the centered 1s interval (timestamps 0.5-1.5s), if two or more vocalization labels were present. 
If only one vocalization label was present,
its label was applied to the whole interval if its duration was 0.2s or greater,
as we observed that children tend to make short vocalizations.
To reduce SD errors, intervals labeled with more than one speaker were discarded. 
To obtain better quality out-of-domain labeled data, non-silent examples with energy below the minimum energy of in-domain CHN vocalizations were discarded. In this way, we intentionally select vocalizations that are close to the CHN and ignore those that have lower energy or can be easily confused with background noise. Note that we didn't apply energy thresholding for in-domain labeled data in order to avoid altering its data distribution. 

For in-domain labeled data partition used in fine-tuning and testing, we followed a leave-one-family-out scheme to ensure training, development and testing sets did not have overlapped families. We divided 70 recordings into four groups based on infant age ranges (1.1-4m, 4-9m, 9-13m, and 13-14m). We randomly selected recordings from one family per age group as the testing set, a small number of recordings as the development set, and the rest of the recordings form the training set. In total, we have 52/6/12 recordings from 15/3/4 families for training/development/testing sets respectively.  

\vspace{-0.2cm}
\subsection{Data augmentation}
\label{sec:data_aug}
Data augmentation has proven beneficial for several speech processing tasks. 
In this study, we used 5 data augmentation techniques implemented using Speechbrain (v0.5.7)~\cite{speechbrain}, as described in~\cite{dawalatabad2021ecapa}, including specAugment~\cite{park2019specaugment}, random chunks of audio dropping, speed pertubation by resampling audio with slightly different sampling rates, reverberation (convolve speech with a room impulse response), noise (add noise to speech with random signal-to-noise ratio ranging from 0-15 dB), and reverberation+noise. We used a room impulse response dataset~\cite{ko2017study} for reverberation corruption. We explored two noise datasets,  MUSAN~\cite{snyder2015musan} and CHiME-Home (CH) ~\cite{foster2015chime}. MUSAN contains 6-hour noises from a variety of sources, such as thunder, paper rustling, fax machine noises, animal noises, etc. CH contains 6.8-hour audio of 4s clips from domestic environment, including child and adult vocalizations, TV and kitchen noises, etc. We also attempted to mine noises from LB and LENA home audio but obtained little improvement, perhaps because W2V2 already learned the distribution of LB/LENA domestic noise.

\section{Experimental Setup}
W2V2 first encodes audio waveforms into latent feature embeddings using several convolutional layers. These latent embeddings are then fed to both a quantizer and a transformer network. The quantizer assigns each latent embedding to a specific learned speech unit from an inventory of quantized units. Some of the latent representations are masked before they are fed into the transformer network. During training, a contrastive loss is applied to transformer outputs and used to predict target audio segment from a context of surrounding audio segments. Detailed pretraining procedures are described in~\cite{wav2vec}. Two versions of W2V2, base (encoded feature size 768 with 12 transformer layers) and large (encoded feature size 1024 with 24 transformer layers), are available, and W2V2 base model is used in our study.

\begin{figure}
  \centering
  \includegraphics[width=1.0\linewidth]{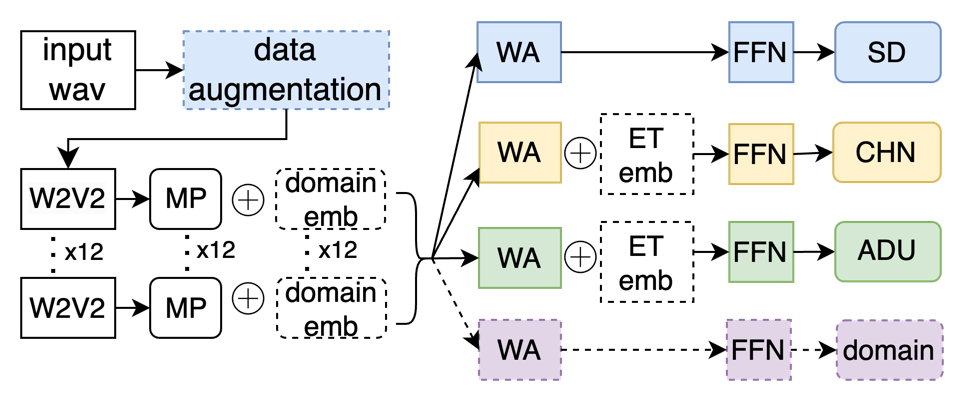}
    \vspace{-0.5cm}
\caption{Overview of model architecture.   W2V2=wav2vec 2.0,
MP=mean pooling, emb=embedding, WA=weighted average, ET=ECAPA-TDNN speaker identity vector, FFN=feed-forward network, SD=speaker
diarization, CHN=child vocalization classifier, ADU=adult vocalization
classifier, domain$\in\{$in,out$\}$ is a boolean marker for out-of-domain data augmentation via embeddings or multi-task learning.  Dashed outline indicates optional modules in some experimental tests.} 
  \label{fig:model}
  \vspace{-0.5cm}
\end{figure}

Figure \ref{fig:model} shows the overall model architecture for fine-tuning W2V2. Three 2-layer feed-forward networks (FFNs) are used as output tiers, including a SD tier and two VC tiers, CHN and ADU (FAN and MAN). The SD tier learns to detect speaker as silent or one of CHN, FAN, MAN, or CXN; if not silent, the corresponding vocalization tier learns to classify the vocalization type. 
We compare W2V2 features extracted from the last transformer layer vs. all 12 transformer layers.
The former applies mean pooling (MP) over time dimension before feeding to FFNs, while the latter applies MP over output $f_{i,t}$ of each transformer layer $i$ for every time step $t$, then a weighted average (WA) layer learns a weight $\alpha_i$ for each output $f_i$, as described in Equation~\ref{eq:WA}: 
\vspace{-0.3cm}
\begin{equation}
\label{eq:WA}
f_{out}=\frac{\sum_{i=1}^{12} \alpha_i ((\sum_{t=1}^Tf_{i,t})/T)}{\sum_{i=1}^{12}\alpha_i}
\end{equation}
We also test applying WA layer first followed by MP but it yields inferior performance.
The loss objective for fine-tuning W2V2 is the average of cross-entropy loss over three output FFNs.
If the training dataset includes out-of-domain data, one of two methods is used to train the network to make any necessary distinction between the processing of in- vs.~out-of-domain data:
concatenating one-hot learnable embeddings to W2V2 features,
or multi-task learning in which a fourth FFN learns a binary domain label.
For the former, we test concatenating one-hot embeddings before or after WA layer and find adding domain embeddings before WA layer gives better results. For the latter, the loss objective is defined as
\vspace{-0.2cm}
\begin{equation}
\label{eq:loss}
L=\alpha_1 L_{SD}+ \alpha_2 L_{CHN} + \alpha_3 L_{ADU} + \alpha_4 L_{domain}
\end{equation}
where $\alpha_1=\alpha_2=\alpha_3=0.33$ and $\alpha_4=0$ if domain FFN is not present, otherwise $\alpha_1=\alpha_2=\alpha_3=0.32$ and $\alpha_4=0.03$.
To further improve overall performance, we explore introducing additional ECAPA-TDNN (ET)~\cite{desplanques2020ecapa} speaker embeddings and data augmentation techniques as described in Section~\ref{sec:data_aug}. 
As ET has shown great performance on adult speaker diarization/recognition tasks, we concatenate ET speaker embeddings pretrained from our labeled data with W2V2 features to provide extra speaker information to potentially improve our model.

%
%
%
%


We pretrain W2V2 on fairseq (v0.12.2)~\cite{ott2019fairseq} using UIUC HAL cluster~\cite{kindratenko2020hal}  with 4 NVIDIA V100 GPUs for 3 weeks until convergence for each experiment. 
We adapt the pretraining recipe of W2V2 base model on 960h Librispeech data with minor changes on minimum (1s) and maximum (10s) audio lengths to save computational memory. We implement fine-tuned model and ET using SpeechBrain. Each fine-tuning experiment is trained 10 epochs with batch size 32 on single NVIDIA GTX 1080 Ti for about 10/40 hours without/with out-of-domain data respectively. With data augmentation, the total training time roughly increases 5 times when there are 5 types of augmentation. We evaluate our model using unweighted F1-scores for each tier over all classes. The epoch with the best average score over three tiers on in-domain development set is used for final evaluation on in-domain testing set. Adam optimizer with learning rates of output FFNs and W2V2 model starting from 1e-4 and 1e-5 respectively is used; scheduler with new-bob technique is used to anneal learning rates based on development set performance after each epoch. ET speaker embedding size is 192. FFN hidden node size is set as 384 (half of W2V2 feature dimension), and one-hot learnable domain embedding size is set as 256. Between the two FFN layers, 1D batch normalization, Leaky Relu activation, and dropout with 0.1 probability are applied sequentially. In total, W2V2 and 3 FFN have 95M and 8M learnable parameters respectively.
Our code and model weights are available.
\footnote{\url{https://huggingface.co/lijialudew/wav2vec_LittleBeats_LENA}} 

\section{Results \& Discussions}
\subsection{Comparisons across different unsupervised pretrained models fine-tuned on in-domain labeled data}
To compare unsupervised pretraining effects, we test 4 unlabeled datasets: \textit{base} (oracle version): 960h unlabeled Librispeech, \textit{Libri960h}: oracle version followed by fine-tuning with labeled 960h Librispeech, \textit{LB1100h}: 1100h LB home audio, and \textit{LL4300h}: 1100h LB+3200h LENA home audio. Each pretrained models was fine-tuned with in-domain labeled LB data. Table \ref{tab:unsupervised_comparisions} presents the results.  W2V2 models pretrained on unlabeled home recordings outperformed original oracle models with/without fine-tuning on Librispeech for all three tiers, and \textit{LL4300h} has the overall best performance. Thus, we use \textit{LL4300h} for the rest of the experiments unless noted otherwise. The results suggest W2V2 effectively learns family audio representation on large-scale unlabeled home recordings across several age groups of infants and toddlers under 5 years old at initial pretraining stage.
For fine-tuning, we compare fine-tuning W2V2 model vs. freezing W2V2 and fine-tuning 3 output FFNs only. We find the former significantly boosts the overall performance compared with the latter. 
We suspect fine-tuning the entire W2V2 makes W2V2 robust to noisy home audio, which often contains clothing rustling, TV noises, toy banging, etc. 
For W2V2 pretrained on \textit{LB1100h}, we obtain comparable performances by using features from either last transformer layer (layer 12) or all 12 layers. For W2V2 pretrained on \textit{LL4300h}, features extracted from all layers benefit VC tiers while features extracted from the last layer help the SD tier. Empirically, features of the top layer encode speaker information and features of lower layers encode vocalization type.  
\begin{table}
  \centering
  \vspace{-0.4cm}
  \setlength{\tabcolsep}{4.0pt}
   \caption{\small F1-scores (\%) for SD, CHN, ADU, and the average of three tiers among W2V2 models pretrained using different unlabeled datasets and fine-tuned on in-domain labeled LB data only, FT: Fine-tune W2V2 and 3 output FFNs/FR:Freeze W2V2 and fine-tune 3 output FFNs} 
\label{tab:unsupervised_comparisions}
      \vspace{-0.2cm}
  \begin{tabular}{lll|cccc}
    \toprule
    model & setting & features & SD & CHN & ADU & Avg\\
    \midrule
    base & FT & all layers & 62.8 & 54.9 & 59.2 & 59.0 \\ 
    Libri960h & FT & all layers &62.5 & 46.5 & 53.5 & 54.2\\ 
\midrule
    LB1100h & FR & layer 12 & 45.5 & 45.3 & 29.6 & 40.1\\
    LB1100h & FR & all layers & 67.9 & 66.2 & 56.4 & 63.5\\
    LB1100h & FT & layer 12 & 70.8 & 66.6 & 60.0 & 65.8 \\
    LB1100h & FT & all layers & 68.2 & 66.3 & 63.0 & 65.8 \\
    \midrule
    LL4300h & FT & layer 12 & \textbf{74.9} & 67.4 & 60.3 & 67.5 \\
    LL4300h & FT & all layers & 71.5 & \textbf{68.9} & \textbf{66.9} & \textbf{69.1} \\
\bottomrule
  \end{tabular}
\label{tab:pretrain}
\vspace{-0.2cm}
\end{table} 
\begin{table}
  \centering
  \setlength{\tabcolsep}{4.0pt}
   \caption{\small F1-scores (\%) trained on both in- and out-of-domain data comparing systems w/o domain tagging, w/one-hot domain embedding, or w/multi-task learning of a domain classifier}
    \label{tab:in_out_domain}
      \vspace{-0.2cm}
  \begin{tabular}{ll|cccc}
    \toprule
    features & domain tagging & SD & CHN & ADU & Avg\\
    \midrule
    layer 12 & - & 68.7 & 68.1 & 67.6 & 68.1 \\
    all layers & - & 68.9 & 69.6 & 70.6 & 69.7 \\
    layer 12 & one-hot & \textbf{70.6} & 70.4 & \textbf{71.9} & \textbf{71.0} \\
    all layers & one-hot & 69.6 & \textbf{70.6} & 68.7 & 69.6 \\
    layer 12 & multi-task & 66.1 & 68.3 & 69.7 & 68.0\\
    all layers & multi-task & 68.4 & 68.6 & 67.3 & 68.1 \\
    \bottomrule
  \end{tabular}
\label{tab:classify}
  \vspace{-0.6cm}
\end{table} 
\vspace{-0.1cm}
\subsection{Effects of adding out-of-domain labeled data}
Table \ref{tab:in_out_domain} shows the results of fine-tuning both in- and out-of-domain labeled data using binary domain learning techniques. Compared with fine-tuning on in-domain data only (see Table \ref{tab:pretrain} last row), we observe that adding out-of-domain data helps improve VC tiers (mostly ADU) but slightly hurts SD tier (see Table \ref{tab:in_out_domain}),
perhaps because out-of-domain microphones cause domain shift in the distribution of speaker diarization cues.
For binary domain learning, one-hot domain embeddings provide marginal benefits while multi-task learning isn't helpful.
\vspace{-0.2cm}
\subsection{Effects of introducing ECAPA-TDNN speaker embeddings and data augmentation}
We pretrain ET using in- only or in- and out-of-domain labeled data, and we achieve unweighted F1-scores of 66.6\% and 67.6\% on labeled LB testing data on SD task respectively. We attempt to concatenate ET embeddings to all three tiers for training but obtain large degradation, which may due to the incompatibility of W2V2 features and ET embeddings learning speaker information simultaneously. Thus, we only concatenate ET embeddings on VC tiers.
Table~\ref{tab:speaker_emb_effects} presents results of ET embedding. We find that ET embeddings are helpful when W2V2 is trained on a limited number of family recordings, such as fine-tuning on \textit{LB1100h} using in-domain labeled data (relative gain 3.3\% for SD, 2.7\% for CHN, and 4.8\% for ADU). To save computational time, we use features from the last transformer layer in \textit{LB1100h}, which performs similarly to using all 12 layers. If W2V2 is trained on a relatively larger number of family recordings, such as fine-tuning on \textit{LL4300h} using all labeled data, ET embeddings provide limited benefits.

We apply 5 types of data augmentation (see Section~\ref{sec:data_aug}), on SD tier only, VC tiers only, or all tiers. Table \ref{tab:noise} summarizes relevant results. We discover that applying data augmentation on SD tier only using MUSAN for additive noise achieves the best performance overall, which shows relative improvement of 12\% on SD tier, 1.7\% on CHN tier, and 1.3\% on ADU tier. 
We observe data augmentation with CH domestic noises degrades VC performances. Probably corrupting home audio with CH child/adult vocalizations injects undesirable background speaker acoustics to target speaker vocalizations.  
When we combine data augmentation with binary domain learning and ET embeddings, we achieved the optimal performance on SD tier with 13.5\% relative improvement.   
\begin{table}
  \centering
\setlength{\tabcolsep}{2.0pt}
   \caption{\small F1-scores (\%) with and w/o ET speaker embeddings}
\label{tab:speaker_emb_effects}
    \vspace{-0.2cm}
  \begin{tabular}{llll|cccc}
    \toprule
    data & model& features & ET emb & SD & CHN & ADU & Avg\\
    \midrule
    In & LB1100h & layer 12 & - & 70.8 & 66.6 & 60.0 & 65.8\\
    In & LB1100h & layer 12 & In & \textbf{73.2} & 68.4 & 62.9 & 68.2\\
    In+Out & LL4300h & all layers & - & 68.9 & \textbf{69.6} & 70.6 & 69.7\\
    In+Out & LL4300h & all layers & In+Out & 70.2 & \textbf{69.6} & \textbf{70.8 }& \textbf{70.2}\\
\bottomrule
\end{tabular}
\label{tab:classify}
  \vspace{-0.3cm}
\end{table} 
\vspace{-0.1cm}
\begin{table}
  \centering
  \setlength{\tabcolsep}{3.0pt}
   \caption{\small F1-scores (\%) using in- and out-of-domain data augmented by SpecAug, reverberation from RIR corpus, additive noise from M:MUSAN/C:CH corpora, and reverberation+noise; features are weighted average of all layers}
    \label{tab:noise}
      \vspace{-0.2cm}
  \begin{tabular}{lll|cccc}
    \toprule
     noise & domain tagging & ET emb & SD & CHN & ADU & Avg\\
    \midrule
    - & - & - & 68.9 & 69.6 & 70.6 & 69.7\\
    C(SD) & - & - & 76.5 & 69.2 & 67.7 & 71.1\\
    M(SD) & - & - & 77.2 & 70.8 & \textbf{71.5} & \textbf{73.2}\\
    M(VC) & - & - & 75.2 & 71.8 & 67.8 & 71.6\\
    M(All) & - & - & 76.2 & \textbf{72.0} & 68.7 & 72.3\\
    \midrule
    M(SD) & one-hot & - & 77.4 & 68.5 & 64.3 & 70.1\\
    M(SD) & one-hot & In+Out & \textbf{78.2} & 70.1 & 68.8 & 72.4\\
\bottomrule
\end{tabular}
\label{tab:classify}
  \vspace{-0.5cm}
\end{table} 
\vspace{-0.2cm}
\section{Conclusions \& Future Work}
This study shows the effectiveness of family audio pretraining and data augmentation to reduce domain mismatch for family audio analysis. 
Pretraining ET speaker embeddings are useful when a limited number of family recordings are used for training. Data augmentation largely helps SD and moderately benefits VC.
In the future, we aim to collect home recordings from more families and explore active learning for quickly adapting current models to new families with minimal labeling effort.


\bibliographystyle{IEEEtran}
\bibliography{mybib}

\end{document}